\documentstyle[11pt,aasms4]{article}
\lefthead{Kepler et al.}
\righthead{Mode identification of Pulsating White Dwarfs using the HST}

\received{31 Jan 2000}
\accepted{17 Mar 2000}

\slugcomment{Submitted to Ap.J.}

\begin{document}

\newcommand{\etal}{et al.}
\newcommand{\msun}{M_\odot}
\newcommand{\mstr}{{\rm M}_{\star}}
\newcommand{\mh}{{\rm M}_{\rm H}}
\title{Mode identification of Pulsating White Dwarfs using the HST
\footnote{Based on observations with the NASA/ESA Hubble Space
Telescope, obtained at the Space Telescope Science Institute, which
is operated by the Association of Universities for Research in
Astronomy, Inc. under NASA contract No. NAS5-26555}}
\author{S.O. Kepler} 
\affil{
Instituto de F\'{\i}sica, Universidade Federal do Rio Grande do Sul, 
91501-970 Porto Alegre, RS -- Brazil} 
\author{E.L. Robinson} 
\affil{
McDonald Observatory and Department of Astronomy, 
The University of Texas, Austin, TX 78712-1083 -- USA}
\author{D. Koester}
\affil{
Institut f\"ur Astronomie und Astrophysik, Universit\"at Kiel,
D--24098 Kiel, Germany}
\author{J.C. Clemens} 
\affil{Department of Physics,
University of North Carolina, Chapel Hill, NC 27599-3255 -- USA}
\author{R.E. Nather} 
\affil{
McDonald Observatory and Department of Astronomy, 
The University of Texas, Austin, TX 78712-1083 -- USA}
\and 
\author{X.~J.~Jiang}
\affil{
Beijing Astronomical Observatory and United Laboratory
of Optical Astronomy, Chinese Academy of Sciences,
Beijing 100101, China}

\begin{abstract}
We have obtained time-resolved ultraviolet spectroscopy
for the pulsating DAV stars G226--29 
and G185--32,
and for the pulsating DBV star PG1351+489
with the Hubble Space Telescope 
Faint Object Spectrograph, to
compare the ultraviolet to the optical pulsation amplitude
and determine the pulsation indices.
We find that for
essentially all observed pulsation modes, the
amplitude rises to the ultraviolet
as the theoretical models predict for  $\ell=1$ 
non-radial g-modes.
We do not find any pulsation mode visible only in the
ultraviolet, 
nor any modes whose phase flips by $180^\circ$ in the ultraviolet,
as would be expected if high $\ell$ pulsations were excited.
We find one periodicity 
in the light curve of
G185--32, at 141~s, which does not
fit theoretical models for the change of amplitude with wavelength
of g-mode pulsations.

\end{abstract}


\section{1. Introduction}

Observations of white dwarf stars are an important probe of
stellar and galactic evolution.  
The properties of individual white dwarfs define the
endpoints for models of stellar evolution,
while the 
white dwarf luminosity function provides an 
observational record of star formation in our galaxy.  
For example, the coolest normal-mass white dwarfs are remnants of 
stars formed in the earliest epoch of star formation, so
their cooling times can tell us the age of the 
galactic disk in the solar neighborhood (Winget et al. 1987, Wood 1992)
and the effects of phase separation and crystallization at extreme
densities (Chabrier et al. 1992, Segretain et al. 1994, Winget et al 1997).

As the number of pulsation modes detected in the pulsating white dwarfs
is insufficient for an inverse solution of the structure of the star,
we must identify the pulsation modes to compare with the theoretical
models and infer the structural parameters.

A crucial step in determining the structure of a white dwarf
from its pulsation periods is to identify the pulsation modes correctly.
The pulsation modes in our models
are indexed with three integers ($k$,$\ell$,$m$)
where $k$ represents the number of nodes in the pulsation eigenfunction
along the radial direction,
$\ell$ is the total
number of node lines on the stellar surface, and $m$ is the number of
node lines passing through the pulsation poles.
Pulsation modes with different indices generally have different 
pulsation periods.
The usual procedure for identifying the mode indices is
(1) calculate theoretical pulsation periods in models of
white dwarfs; (2) compare the pattern of theoretical periods
to the observed pattern of periods; (3) adjust the models to
bring the theoretical and observed patterns into closer agreement.
The problems with this procedure are clear:
it does not work for white dwarfs with only a few excited pulsation modes,
as it places too few constraints on the stellar structure;
and, given the complexity and sophistication of the theoretical
calculations and the large number of possible pulsation modes, 
there is ample opportunity to misidentify modes.
Other methods of mode identification must be used to avoid these
problems.

\section{Mode Identification Using Time-Resolved UV Spectroscopy}

Time-resolved ultraviolet spectroscopy provides an independent
method for determining the pulsation indices of white dwarfs.
The amplitudes of {\it g}-mode pulsations depend strongly
on $\ell$ at wavelengths shorter than 3000~\AA.  
Figure 1 shows how the amplitude depends on wavelength and $\ell$
for the lowest-order modes of  pulsating white dwarfs.
The amplitude of all modes increases towards the ultraviolet
but the amplitude increases more for $\ell = 2$ than for $\ell = 1$.
The differences are even greater for modes with higher $\ell$. Note
the predicted $180^\circ$ phase flip of the $\ell=4$ mode for the
DAV models at wavelengths shorter than 1500\AA, indicated by
the negative amplitudes.

The increase in amplitude from optical to ultraviolet wavelengths is caused
by two effects: the increasing effect of the temperature on the flux,
and the increasing effect of limb darkening in the ultraviolet.
The differences among the amplitudes of modes with different $\ell$ 
are caused mainly by limb darkening.
The brightness variations of non-radially pulsating white dwarfs 
are due entirely to variations in effective temperature;
geometric variations are negligible (Robinson, Kepler, \& Nather 1982). 
The normal modes divide the stellar surface into zones of
higher and lower effective temperature 
that can be described by spherical
harmonics; modes of higher $\ell$ have
more zones than those of lower $\ell$.  
From a distance, we can measure only the integrated surface
brightness, which includes the effects of limb darkening,
so modes of high $\ell$ are normally
washed out by the cancellation of different zones.
But at ultraviolet wavelengths, 
the effects of limb darkening increase drastically, 
decreasing the contribution of zones near the limb.  
Consequently, modes of higher $\ell$ are cancelled less 
effectively in the UV and their amplitudes increase
more steeply at short wavelengths than those of low $\ell$.
Theoretical calculations of the amplitudes require good
model atmospheres but are entirely independent of the
details of pulsation theory and white dwarf structure calculations.

Robinson et al.~(1995) used this method to determine 
$\ell$ for the pulsating DA white dwarf G117--B15A.
They measured the amplitude of its 215~s pulsation in the ultraviolet
with the HST high-speed photometer and identified it as an $\ell=1$ mode.
With the correct value of $\ell$, they found that the mass
of the surface hydrogen layer in G117--B15A was between
$1.0 \times 10^{-6}$ and $8 \times 10^{-5}\ M_\odot$, 
too thick to be consistent with models invoking thin hydrogen
layers to explain the spectral evolution of white dwarfs.
They also found $T_{\mathrm{eff}} = 12\,375 \pm 125$~K,
substantially lower than the accepted temperature at that time,
but close to the presently accepted temperature
(Koester et al. 1994, Bergeron et al. 1995, Koester \& Allard 2000).

To extend these results,
we observed the pulsating DA white dwarfs G226--29 (DN Dra) and
G185--32 (PY Vul), and  the DBV PG1351+489 (EM UMa)
with
10 sec/exposure RAPID mode of the (now decomissioned)
Faint Object Spectrograph (FOS) of the
Hubble Space Telescope.
We used the blue Digicon detector and the G160L grating over the
spectral region 1150 \AA\ to 2510 \AA.

\section{Observations}
\subsection{G226--29}
G226--29, also called DN Dra, LP~$101-148$, and WD~1647+591,
is the brightest known pulsating
DA white dwarf (DAV or ZZ Ceti star), with $m_{v}=12.22$.
At a distance of just over 12~pc, it is the closest ZZ~Ceti star
(optical parallax of $82.7\pm 4.6$ mas,
Harrington {\&} Dahn 1980;
Hipparcos parallax of $91.1\pm 2.1$ mas, Vauclair et al. 1997).
Its pulsations were discovered by
McGraw \& Fontaine (1980),
using a photoelectric photometer attached to the MMT telescope.
They found a periodicity at 109~s with a 6 mma (mili modulation
amplitude) amplitude near 4200~\AA.
Kepler, Robinson, {\&} Nather (1983) used
time-series photometry to solve the
light curve and interpret the variations as an equally spaced triplet
with periods near 109~s.
The outer peaks have similar  amplitudes, near 3~mma,
and are separated by a frequency
$\delta f=16.14 \,\mu$Hz from the central peak,
which has an amplitude of 1.7~mma.
These results were confirmed by Kepler et al. (1995a), using
the Whole Earth Telescope, who also showed that no other pulsation were present
with amplitudes larger than 0.4~mma.
G226--29 has the simplest mode structure, the second smallest overall pulsation
amplitude, and the shortest dominant period of any pulsating white dwarf.


For G226--29, the very short (109~s) period triplet leads to a seismological 
interpretation of the structure by Fontaine {\etal} (1994),
who show the star should have a thick hydrogen layer
($\log q = \log\mh /\mstr = -4.4 \pm 0.2$)
if the observed triplet is the 
rotationally split $\ell=1$, $k=1$ mode.
Kepler {\etal} (1995a), assuming an $\ell =1$, $k=1$ triplet, 
also derived an hydrogen layer
mass about $10^{-4} \mstr$. Higher $k$ values would imply an unreasonably
thick hydrogen layer.

Several recent spectroscopic studies show G226--29 to be one of the hottest
of the ZZ~Ceti stars, suggesting that we may be observing it as it 
enters the instability strip.
The absolute effective temperature of this star is not settled, because one can
derive two different effective temperatures for a given gravity using
optical spectra, and also because there are  
uncertainties about the best convective efficiency to use in model atmospheres
(Bergeron et al. 1992, 1995, Koester \& Vauclair 1997).
Fontaine et al. (1992) derive $T_{\mathrm{eff}}=13,630 \pm 200$ K and 
$\log g = 8.18 \pm 0.05$, 
corresponding to a stellar mass of $0.70 \pm 0.03\msun$,
based on high signal-to-noise optical spectra and ML2/$\alpha=1$
 model atmospheres.
This effective temperature places G226--29 near the blue edge of their
ZZ~Ceti instability strip.
Kepler \& Nelan (1993) used published IUE spectra and optical
photometry to derive $T_{\mathrm{eff}}=12\,120 \pm 11$ K, assuming $\log g=8.0$;
their ZZ~Ceti instability strip is much cooler,
12\,640~K $\geq T_{\mathrm{eff}} \geq $ 11\,740~K.

Koester {\&} Allard (1993) use the Lyman~$\alpha$ line profile
to derive a parallax-consistent solution of $T_{\mathrm{eff}}=12\,040$~K and
$\log g =8.12$.
Bergeron et al. (1995) found $T_{\mathrm{eff}} = 12\,460$, and $\log\, g = 8.29$
for an $\alpha = 0.6$ ML2 model which fits the IUE and optical spectra
simultaneously;
their instability strip spans $12\,460 \geq T_{\mathrm{eff}} \geq 11\,160$ K,
placing G226--29 on the blue edge.
Koester, Allard \& Vauclair (1995) show G226--29 must
have nearly the same temperature as L~19--2 and G117--B15A, at about 12\,400~K,
in agreement with Kepler {\&} Nelan (1993), Koester {\&} Allard (1993),
and Bergeron et al. (1995).
Kepler et al. (1995b) found $T_{\mathrm{eff}} = 13\,000 \pm 110$, and $\log\, g = 8.19 \pm
0.02$, which corresponds to a mass of $0.73\, \msun$, from the optical
spectra alone, using Bergeron's ML2 model atmosphere.
Giovannini et al. (1998), using the same optical spectra as Kepler et al.,
but using Koester's ML2 model atmosphere, obtained
$T_{\mathrm{eff}} = 13\,560 \pm 170$, and $\log\, g = 8.09 \pm 0.07$,
which corresponds to a mass of $0.66\, \msun$, for a DA evolutionary model
of Wood (1995). Koester \& Allard (2000) obtained 
$T_{\mathrm{eff}}=12\,050 \pm 160$~K,
and $\log g=8.19 \pm 0.13$, using IUE spectra, V magnitude and parallax.
This general agreement on the value of
$\log g$ suggests the mass is around $0.70\msun$.
The effective temperature is 
most probably 12\,100~K,
consistent with the IUE continuum, parallax and optical line profiles
simultaneously.

G226--29 was observed with the HST six times, each time for 3 hours,
between September 1994 and December 1995.
As the star is bright and fairly hot,
the time-averaged spectrum from
the total of 18.6 hrs of observation has a high signal-to-noise ratio
(Figure \ref{Fig3}).

\subsection{G185--32}
The largest-amplitude pulsations of G185--32 
have periods of 71~s, 141~s and 215~s (McGraw et al. 1981).
We observed G185--32 with HST for a total of 7.1~hr on 31 Jul 1995.
The Fourier transform of the UV and Zeroth order (see section \ref{zero}) 
light curve (Figure \ref{g185panel}) shows the periods we have
identified for this star.

\subsection{PG1351+489}

PG1351+489 is the DBV with the simplest pulsation spectrum, and
therefore the one which requires the shortest data set to measure
its amplitude. Its pulsations, discovered by
Winget, Nather \& Hill (1987), have a
dominant period  at 489~s and a peak-to-peak blue amplitude near 0.16~mag.
The light curve also shows the first and second harmonics of the
this period ($f_0$), plus
peaks at 1.47~$f_0$, 2.47~$f_0$ and 3.47~$f_0$, with lower amplitudes.
We observed PG1351+489 for 4 consecutive orbits of HST,
for a total of 2.67~hr.
The ultraviolet and Zeroth order (see section \ref{zero}) Fourier spectra 
(Figure \ref{pg1351dft})
show only
the 489~s period and its harmonic at 245~s above the noise, and a
possible period at 599~s.

\section{Zeroth Order Data\label{zero}}

Although not much advertised by the STScI, 
the zeroth order (undiffracted)
light from an object falls onto the FOS detector when using the G160L
grating and
provides simultaneous
photometry of the object with an effective wavelength around 
3400~\AA~(see Figure \ref{Fig9})
(Eracleous \& Horne 
1996).\footnote{The data 
can be extracted from pixels 620 to 645 from the c4 files.}
The simultaneous photometry from the zeroth
order light was crucial to the success of this project.
As the zeroth order light has a counting rate
around 100 times larger than the total light collected in the first order
time resolved spectra, 
it can also be used to search for low amplitude
pulsations. In the searched range of 800~s to
20~s, no new ones were found for any star observed
in this project, to a limit around 8~mma.

The calibration pipeline of the HST data contains a transmission
curve for the zeroth order data measured on
the ground prior to launch
(Figure \ref{Fig10}), but our data are inconsistent with
this transmission curve. We will discuss this later in section \ref{amplitudes}.

\section{Data Set Problems}
We detected two significant problems in the FOS data sets
on G226--29, the first star we observed.
First, we found a $\sim 3$\% modulation of the 
total count rate on a time scale similar to the HST orbital period
(see Figure \ref{Fig2}).
This modulation is probably caused by a combination of factors.
We used a 1 arcsec entrance aperture and 
a triple peak-up procedure to center the star in the aperture
for the first 5 observations.
The triple peak-up process
yields a centering accuracy of only $\pm 0.2$
arcsec which, when coupled with the 0.8 arcsec
PSF of the image, produces light loss at the 
edges of the aperture of at least
a few percent.
As the position of the star image in the aperture wanders 
during the HST orbit,
the amount of light lost at the aperture varies, modulating
the detected flux.

The second problem became evident
when we compared the observed spectrum to the spectrum of
G226--29 obtained with
IUE and to model atmospheres for DA white dwarfs
(Koester, Allard \& Vauclair 1994). We found a spurious ``bump''
in the FOS spectrum in a 75~\AA\ region just to the blue of 1500~\AA.
The bump is not subtle: it rises 25\% above the surrounding
continuum (see Kepler, Robinson \& Nather 1995).

The excess was caused by a scratch on the cathode of the FOS 
blue detector, in a region used only for the G160L grating, 
for which the pipeline flat field did 
not correct properly.
The scratch is at an angle with respect to the diode array
so that the wavelength of the bump in the spectrum changes
as the position of the spectrum on the cathode changes.
The pipeline flat field was obtained with a 0.04 arcsec 
centering accuracy in the 4.3 arcsec aperture
and is not accurate for any other aperture or position. 
For our sixth and last
observation, we used the upper 1.0 pair aperture to minimize 
the flat fielding problem. 
A method for re-calibrating all post-COSTAR observations with the
G160L has recently been devised,
including an Average Inverse Sensitivity 
correction.\footnote{The AIS should be used with STSDAS task calfos,
using the FLX\_CORR omit option.}
After re-calibration, there was significant improvement
in our data, but there is still some problem with the flux calibration
redwards of 2200\AA, as well as some scattered light into the
Ly$\alpha$ core.

To identify the pulsation modes in a white dwarf we
need to know only the fractional amplitudes of the pulsations as
a function of wavelength.
As the fractional amplitudes are immune to multiplicative
errors in the calibration of the spectrograms, and as
both problems we found are multiplicative,
these problems do not affect our results.
Data for wavelengths shorter than 1400~\AA~have known
scattered light correction problems which, being additive,
reduce the accuracy of our measured amplitudes
by an uncertain amount.

\section{Models}

The model atmospheres used to fit the time-averaged spectra, and
to calculate the intensities at different
angles with the surface normal, 
were calculated with a code written by Koester
(Finley, Koester, Basri 1997).
The code uses the ML2/$\alpha$=0.6 version of the standard mixing length
theory of convection and includes the latest version of the quasi-molecular
Lyman $\alpha$ opacity after Allard et al. (1994).
This choice of convective efficiency allows for a consistent
temperature determination from  optical and ultraviolet
time-average spectra (Bergeron et al. 1995,
Vauclair et al. 1997, Koester \& Allard 2000).
ML2/$\alpha=0.6$ is also consistent with
the wavelength dependence of the amplitude 
observed in G117--B15A by Robinson et al. (1995),
according to Fontaine et al. (1996).

To calculate the amplitudes of the pulsations, we require the specific
intensities emitted by the white dwarf atmospheres as a function
of wavelength, emitted angle, effective temperature, gravity, and
chemical composition.

Robinson, Kepler \& Nather (1982) calculated the
luminosity variations of g-mode pulsations, and Robinson et al. (1995)
expanded the results to include explicitly an arbitrary
limb darkening law $h_\lambda(\mu)$, where $\mu = \cos\theta$.
If we call the coordinates in the frame of pulsation  $(r,\Theta,\Phi)$,
the coordinates in the observer's frame $(r,\theta,\phi)$,
and assume 
\[r = R_o( 1 + \epsilon\, \xi_r)\]
with \[\xi_r = {\rm Real} \{Y_{\ell m}(\Theta, \Phi) e^{i\sigma t}\}\]
and assume low amplitude adiabatic pulsations
\[\frac{\delta T}{T} = \nabla_{ad}\frac{\delta P}{P},\]
then the amplitude of pulsation at wavelength $\lambda$, $A(\lambda)$,
defined as 
\[A(\lambda)\cos \sigma t = \frac{\Delta F(\lambda)}{F(\lambda)}\]
is given by
\begin{equation}
A(\lambda) = \epsilon \, Y_{\ell m}(\Theta_0,0)
\left(\frac{1}{I_{0\lambda}}\frac{\partial I_{0\lambda}}{\partial T}\right)
\left(R_0 \frac{\delta T}{\delta r}\right)
\times \frac{\mbox{$\normalsize \int h_\lambda(\mu)P_\ell(\mu)\mu d\mu$}}
{\mbox{$\normalsize \int h_\lambda(\mu )\mu d\mu$}}
\end{equation}
where $I_{0\lambda}$ is the sub-observer intensity, 
i.e., the intensity for $\cos \theta=1$,
and
$P_\ell(\mu)$ are the Legendre polynomials.
We have defined $(\Theta_0,0)$
as the coordinates of the observer's $\theta=0$ axis with respect
to the $(r,\Theta,\Phi)$ coordinate system.

By taking the ratio $A(\lambda)/A(\lambda_0)$,
we can eliminate
the perturbation amplitude $\epsilon$,
the effect of the inclination 
between the observer's line of sight and the pulsation axis 
[$Y_{\ell m}(\Theta_0,0)$],
and the term $R_0(\delta T/\delta r)$, 
all of which cancel out. The term 
$\left(\frac{1}{I_{0\lambda}}\frac{\partial I_{0\lambda}}{\partial T}\right)$
and the limb darkening function $h_\lambda(\mu)$ must be calculated
by the model atmosphere code, and the amplitude ratio is then
calculated by numerical integration. For g-mode pulsations, the amplitude
ratio is therefore a function of $\ell$.

\section{Fit to the Time Averaged Spectra\label{time}}

For G226--29, we 
used our high S/N average spectra to fit to Koester's model
atmospheres, constrained by HIPPARCOS
parallax (Vauclair et al. 1997) to  obtain $T_{\rm{eff}} = 12\,000 \pm 125$~K,
$\log g = 8.23 \pm 0.05$.
The pure spectral fitting does not significantly constrain $\log g$, but
confines $T_{\rm{eff}}$ to
a narrow range. The parallax, on the other hand, very
strongly constrains the luminosity, and (via the mass-radius relation
and $T_{\rm{eff}}$) also the radius, and
thus $\log g$ (Figure \ref{koester}).

For G185--32, Koester \& Allard (2000) also show that the V magnitude
and the parallax can be used to constraint the gravity and they obtained
$\log g=7.92 \pm 0.1$ and $T_{\mathrm{eff}}=11\,820 \pm 110$~K.
Our time-averaged HST spectrum (Figure \ref{g185s})
gives an effective temperature
value of $T_{\mathrm{eff}}=11\,770 \pm 30$~K, for such surface gravity.
The time-averaged HST spectra alone cannot, for any star studied here, 
constrain both
the effective temperature and the surface gravity simultaneously,
and that is the main reason the parallax is used when available, as
a further constraint. 

For PG1351+489, whose spectra is shown in Figures \ref{pg1351s} and 
\ref{pg1351s1},
we don't have a parallax measurement, and
pure He models give similar fit for 
$\log g = 7.50$ and $T_{\mathrm{eff}}=24\,090 \pm 620$~K, or
$\log g = 7.75$ and $T_{\mathrm{eff}}=24\,000 \pm 210$~K, or
$\log g = 8.00$ and $T_{\mathrm{eff}}=23\,929 \pm 610$~K.
Our quoted values are for pure helium atmosphere, but to complicate
things further, Beauchamp et al (1999)
can fit the optical spectra with
$T_{\mathrm{eff}} = 26\,100~{\mathrm{K}}$, $\log g=7.89$ for
pure helium model, but 
$T_{\mathrm{eff}} = 22\,600~{\mathrm{K}}$, $\log g=7.90$
allowing some hydrogen, undetectable in the optical spectra.

\section{Optical Data for G226--29}

Since most of the ground-based
time-series observations to date on G226--29 were
obtained in white light with biakali photocathode detectors, 
we obtained simultaneous UBVR time
series photometry using the Stiening photometer (Robinson et al. 1995)
attached to the 2.1~m Struve telescope at McDonald Observatory,
in March to June 1995, for a total on 39.2 hr of 1~sec exposures
on the star.
We then transformed the time base
to Barycentric Julian Dynamical Time (BJDD) to eliminate the phase
shift introduced by the motion of the Earth relative to the
barycenter of the solar system. We calculated a Fourier transform
of the intensity versus time, for
each of the UBVR colors, and measured the amplitudes and phases
for the three modes,  called $P_0$, $P_1$, and $P_2$,
the triplet around 109~s (Table \ref{UBVR}).
Even though we use this nomenclature, we are {\it not} assuming the
three modes correspond to an $\ell=1$ mode split by rotation,
as we are studying the $\ell$ value for each component independently.
Mode  $P_0$ has a period of 109.27929~s,
$P_1$ has a period of 109.08684~s,
$P_2$ has a period of 109.47242~s. The phases of the three modes
are the same in all filters, as expected for $g$-mode pulsations
(Robinson, Kepler \& Nather 1982).

As the HST data on G226--29 were spread out over 16 months, the ephemeris 
of our previous optical data were not accurate enough
to bridge the resulting time gaps,
requiring us to obtain an additional optical  data set 
to improve the accuracy of the pulsation ephemeris.
We observed the star again with the McDonald Observatory 2.1~m telescope for 
1.7 hr from 8--15 May 96,
1.4 hr on 7 Feb 97,
2.7 hr on 6 May 1997, and 13.6 hr from 3 Jun 1997 to 11 Jun 1997 using the
85cm telescope at Beijing Astronomical Observatory with a Texas 3 star
photometer.  
With this data set we were able to improve the ephemeris
for the three pulsations enough to cover the HST data set and, using the 1995
data set, to extend back to the 1992 Whole Earth Telescope data set. Our
new ephemeris, accurate from 1992 to 1997, is:
\[P_0 = 109.279\,299\,45 \,{\rm sec} \pm 3.3 \times 10^{-6} \,{\rm sec},\]
\[T_{\rm max}^0(\rm BJDD) = 244\,8678.789\,330\,34 \pm 3.7 \,{\rm sec};\]
\[P_1 = 109.086\,874\,54 \,{\rm sec} \pm 1.8 \times 10^{-6} \,{\rm sec},\]
\[T_{\rm max}^1(\rm BJDD) = 244\,8678.789\,951\,19 \pm 2.0 \,{\rm sec};\]
\[P_2 = 109.472\,385\,02 \,{\rm sec} \pm 6.5 \times 10^{-7} \,{\rm sec},\] 
\[T_{\rm max}^2(\rm BJDD) = 244\,8678.789\,541\,96 \pm 0.5 \,{\rm sec}.\]

Our data set is not extensive enough to extend the ephemeris back to the 
1980--1982 discovery data.

\section{Ultraviolet Amplitudes\label{amplitudes}}

To analyze the HST data for the pulsation time variability,
we first integrated the observed spectra into one bin, 
by summing over all wavelengths,
to obtain the highest
signal to noise ratio. We then transformed the time base
to Barycentric Julian Dynamical Time (BJDD), 
and calculated the Fourier transform
of the intensity versus time.
For all three stars we conclude that the ultraviolet (HST) data sets showed
only the pulsation modes previously detected at optical wavelengths.

\subsection{G226--29}
Figure \ref{Fig4} shows the Fourier spectra
of the light curve of G226--29, converted to amplitude, 
and the effects of subtracting,
in succession, the three pulsations we have detected. 
The periods used in the subtraction are those of our new ephemeris,
but the phases and amplitudes were calculated with a linear
least squares fit to the HST data by itself.
The residual
after this process is probably due to imperfect ``pre-whitening''
and does not indicate the presence of other pulsations.

The complex spectral windows arises from the fact that the beat
period between the pulsations is around 17 hr, and the length
of each run of HST was about 3 hr. We can identify
the two largest modes, 
marked $P_1$ and $P_2$ in Figure \ref{Fig4},
at 109.08 and 109.47~s, but the central
peak, which has a smaller amplitude, is largely hidden in the complex
spectral window. The Fourier transform shows that we cannot 
totally separate the smallest amplitude pulsation, called $P_0$,
from the two largest amplitude pulsations, called $P_1$ and $P_2$
modes; its amplitude and phase have large uncertainties in the
HST data set. Note that we use the periods measured in the optical
to subtract the light curves (prewhitening), as they are much
more accurate than the HST values. The fact that the subtraction
works confirms that the optical periods are the same as the ones 
found in the UV.

After concluding the the ultraviolet (HST) data sets presents
only the pulsation modes previously detected, 
we integrated the observed spectra in 50~\AA\ bins.
After determining that the pulsations at all wavelengths were in phase,
we fitted three sinusoids simultaneously to each wavelength bin,
with phases fixed to the values obtained from the co-added spectra.
Figure \ref{Fig5} shows the measured amplitudes. 
We then normalized the amplitude of the pulsations by their
amplitude at 5500~\AA~to compare with the theoretical models.
We used the zeroth order data set to
check the amplitude at U,
which has similar wavelength, and noticed that, unlike any other
observation of the star, the ratio of amplitudes between modes
$P_1$ and $P_2$ changed significantly, making the use of the 
UBVR measurements unreliable, as they were not simultaneous with the
HST data. We therefore renormalized the optical data using the amplitude
ratios derived from the zeroth order data,
assuming its effective wavelength is 3400~\AA~$\pm$~100~\AA.
The published transmission curve for the zeroth order data,
convolved with the models,
would demand an amplitude of
pulsation much larger than observed.
The ultraviolet
efficiency of the mirror must be much lower than measured on the
ground, but the effective wavelength is consistent with our
measurements, to within our uncertainty of around 100~\AA.
Even though the central wavelength of the mirror
is uncertain, the amplitude vs. wavelength changes only by
a few percent over 100~\AA, so we include the uncertainty in
the wavelength as an uncertainty in the normalization.
We note that the observed ultraviolet amplitudes were used
to test this effective wavelength and not only are they consistent with it,
they exclude any effective wavelength shorter than 3100~\AA,
as it would produce a much higher amplitude than observed.
Note that to calculate
$A(\lambda)/A(\lambda_{\mathrm{ref}})$ 
we only need one amplitude, $A(\lambda_{\mathrm{ref}})$,
for normalization [see equation (1)]. By using the Zeroth order amplitude,
we do not need the UBVR amplitudes.

\begin{figure} 
\plotone{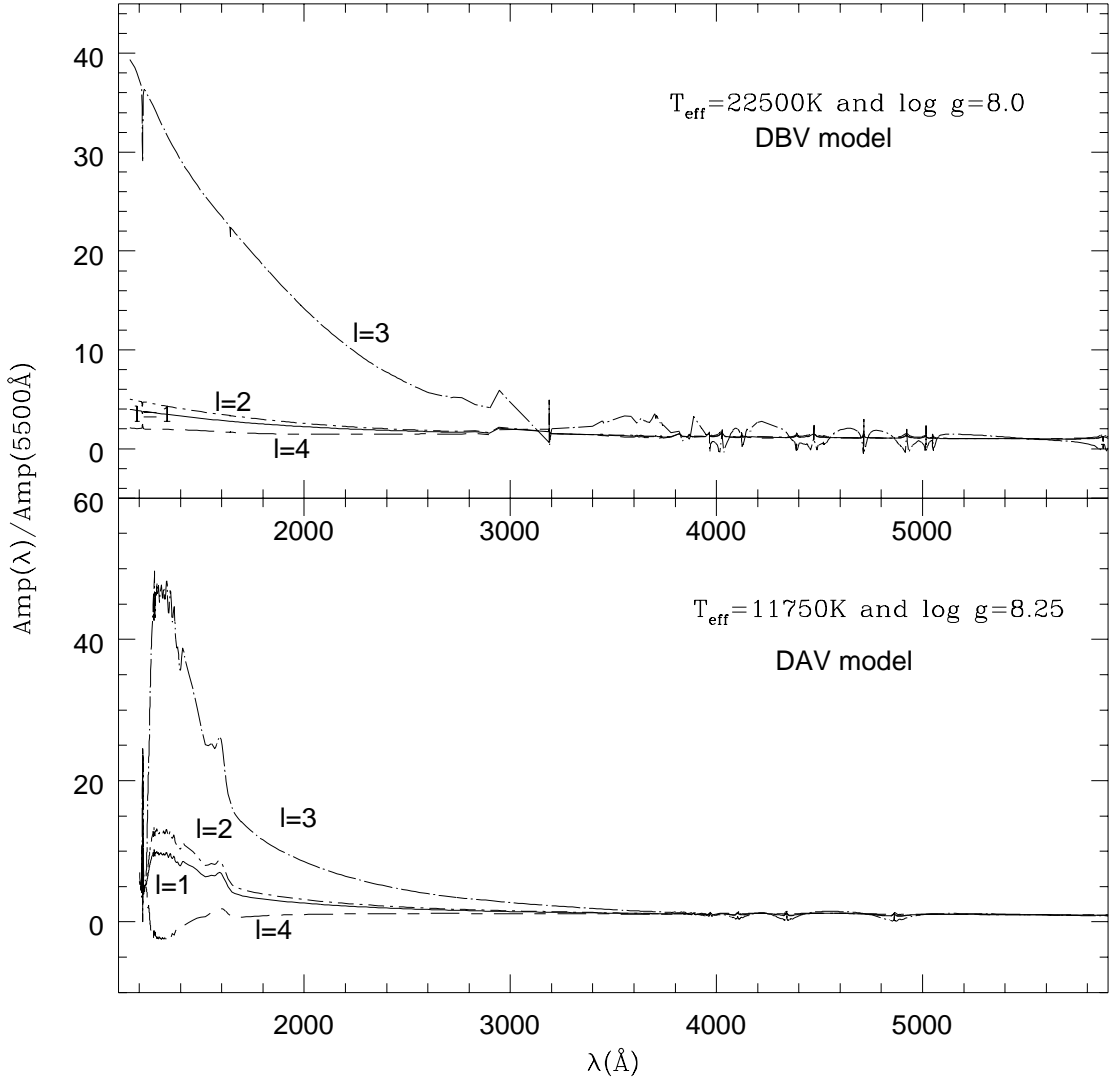}
\figcaption[fig1.ps]{The amplitudes of the $\ell = 1$ to $\ell = 4$ pulsation
modes as a function of wavelength for DA and DB models.
These amplitudes, for different values of $T_{\mathrm{eff}}$
and $\log g$
were calculated from the most recent version of the
model atmosphere code by Koester (Finley, Koester \& Basri 1997).
\label{Fig1}}
\end{figure}

\begin{figure} 
\plotone{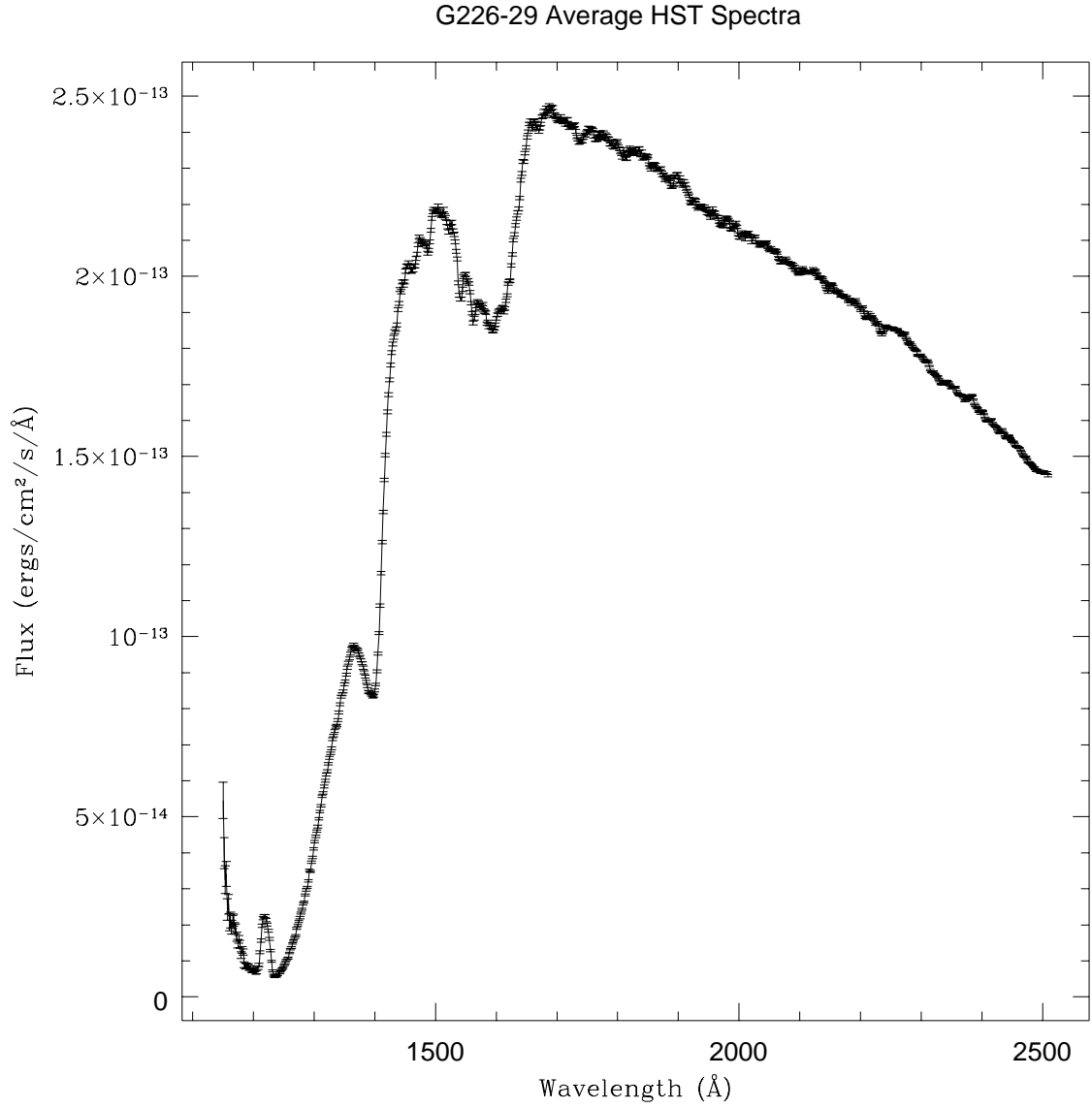}
\figcaption[fig2.ps]{Average FOS spectrum of the pulsating DA white dwarf G226-29,
after re-calibration.
\label{Fig3}}
\end{figure}

\begin{figure} 
\epsscale{0.8}
\plotone{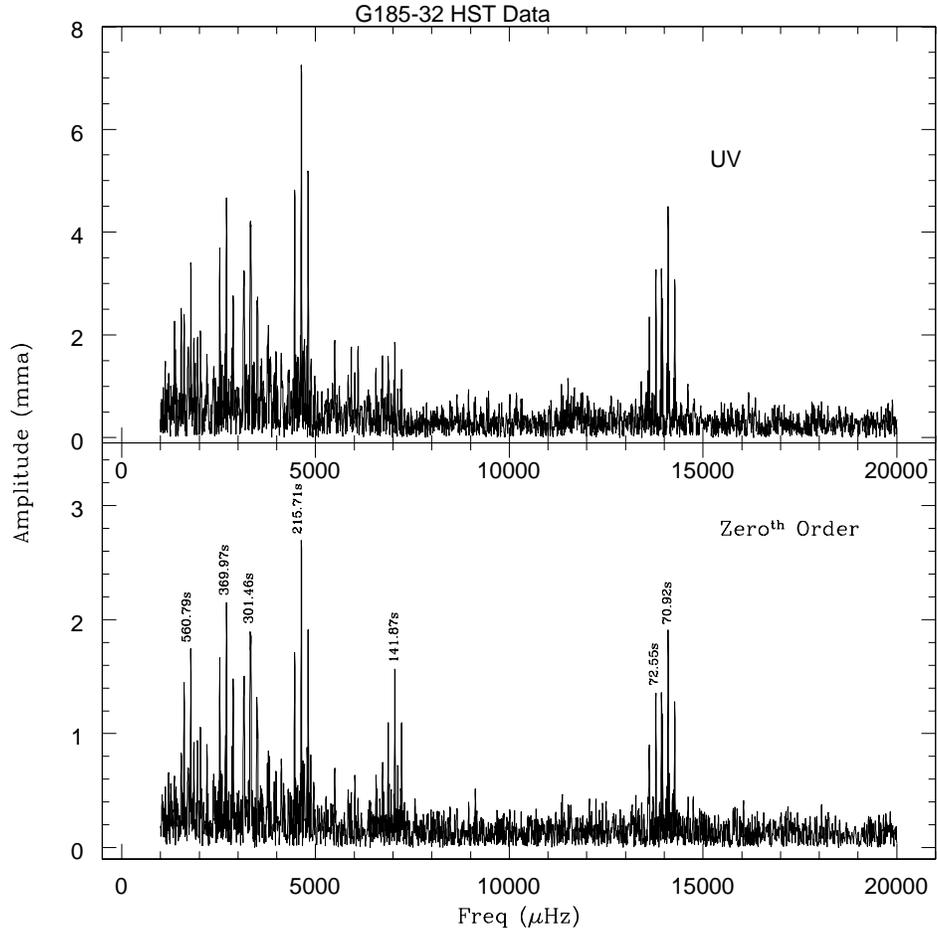}
\figcaption[fig3.ps]{Fourier transform of the UV (upper panel) and
Zeroth Order
(lower panel) data for G185--32.
The peaks not annotated are artifacts introduced by gaps
in the data (aliasing).
\label{g185panel}}
\end{figure}

\begin{figure} 
\epsscale{0.8}
\plotone{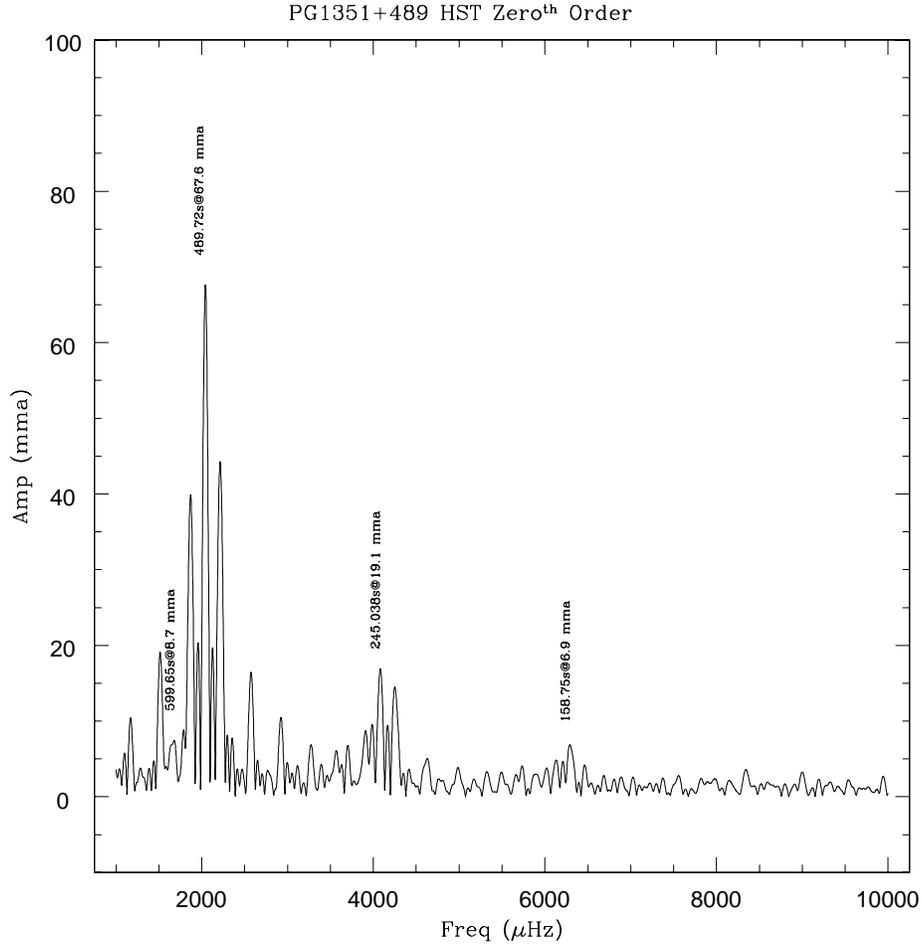}
\figcaption[fig4.ps]{Fourier transform of the
Zeroth Order data for PG1351+489
showing the 489~s and 245~s periodicities. The peak at 158.75~s
is only marginally significant. The peak at 599.65~s
has an amplitude of 8.7~mma on the Zeroth Order data,
which corresponds to $4.3~\langle amp \rangle$ and therefore is
significant.
\label{pg1351dft}}
\end{figure}

\begin{figure} 
\plotone{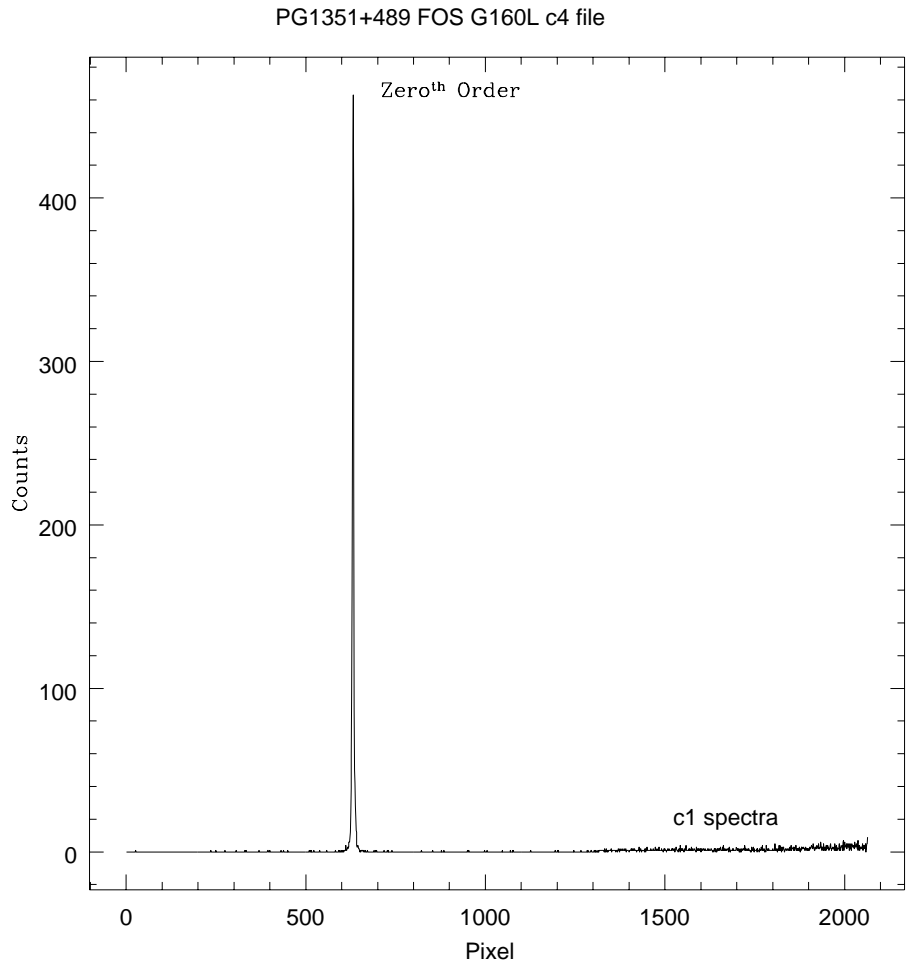}
\figcaption[fig5.ps]{Placement and count rate of the Zeroth
order data,
the undiffracted image of the target object.
\label{Fig9}}
\end{figure}

\begin{figure} 
\plotone{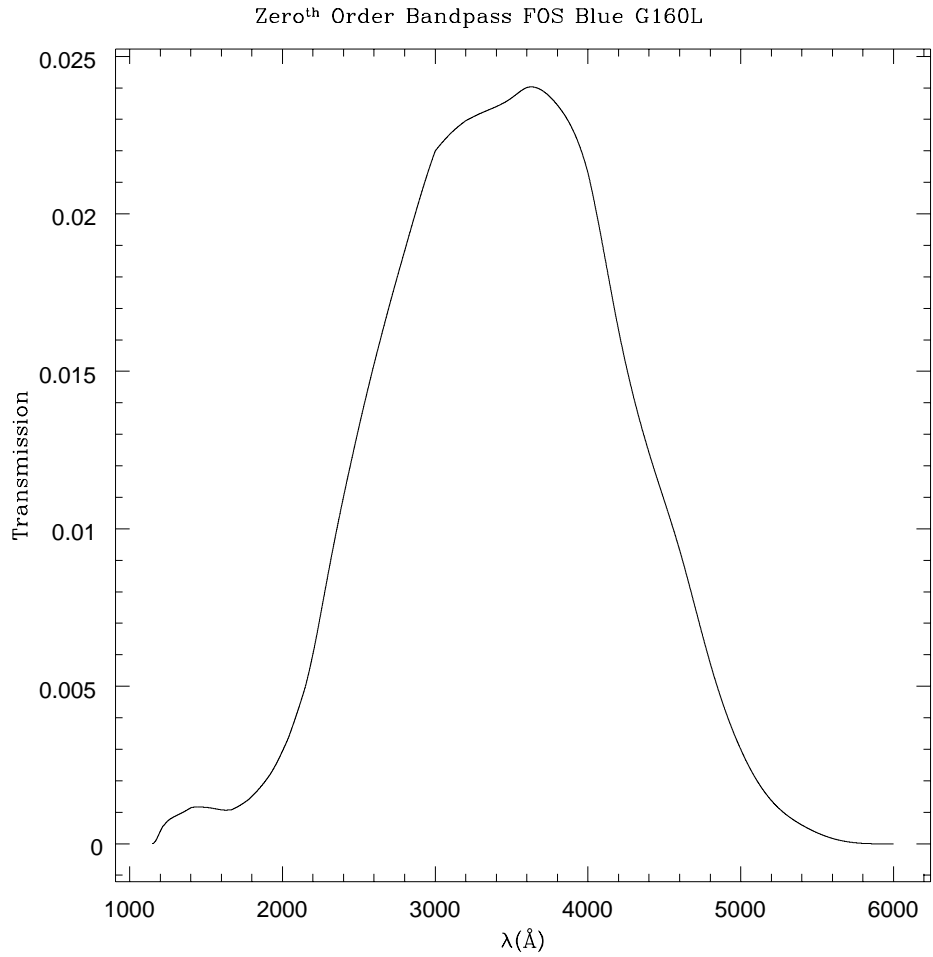}
\figcaption[fig6.ps]{Band pass for the Zeroth 
data as measured on the ground prior to installation of the FOS on HST.
\label{Fig10}}
\end{figure}

\begin{figure} 
\plotone{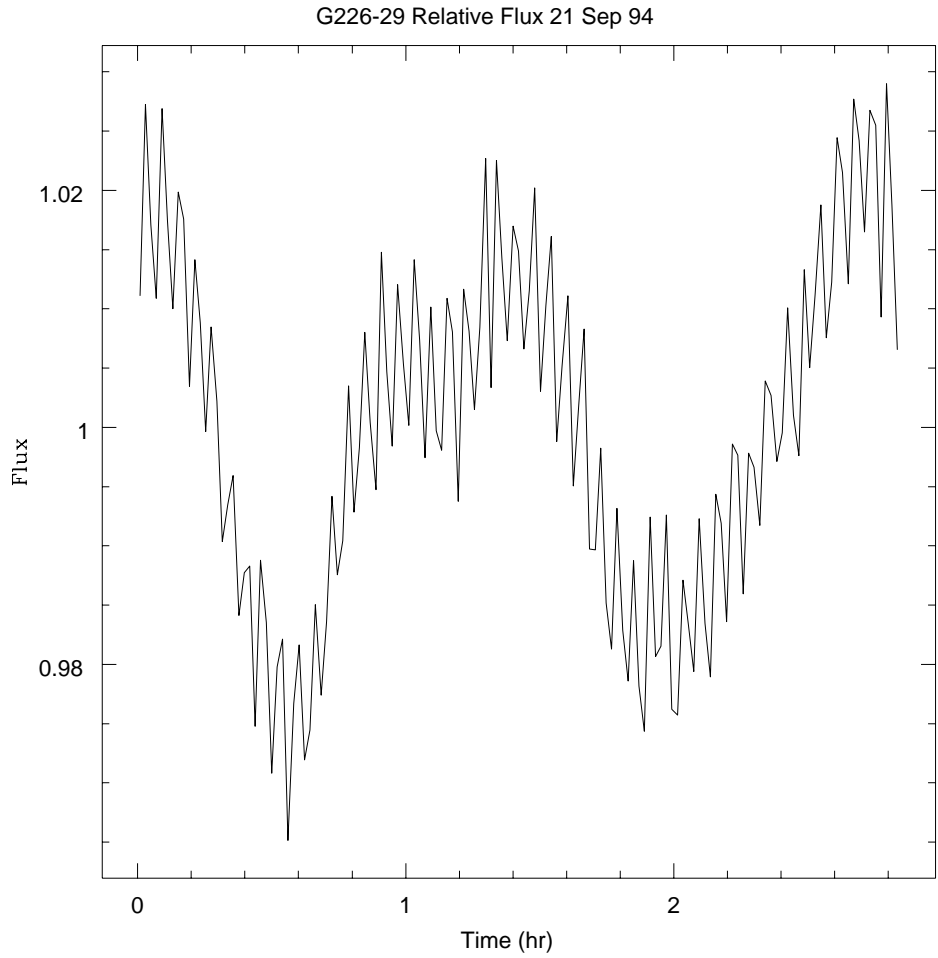}
\figcaption[fig7.ps]{Light loss with the 1.0 arcsec aperture. In this figure
we plot the count rate summed through all wavelengths, divided by
the average, versus time. The rapid variation on
a time scale of 100 sec is caused by the
pulsations of the star.
\label{Fig2}}
\end{figure}

\begin{figure} 
\epsscale{0.7}
\plotone{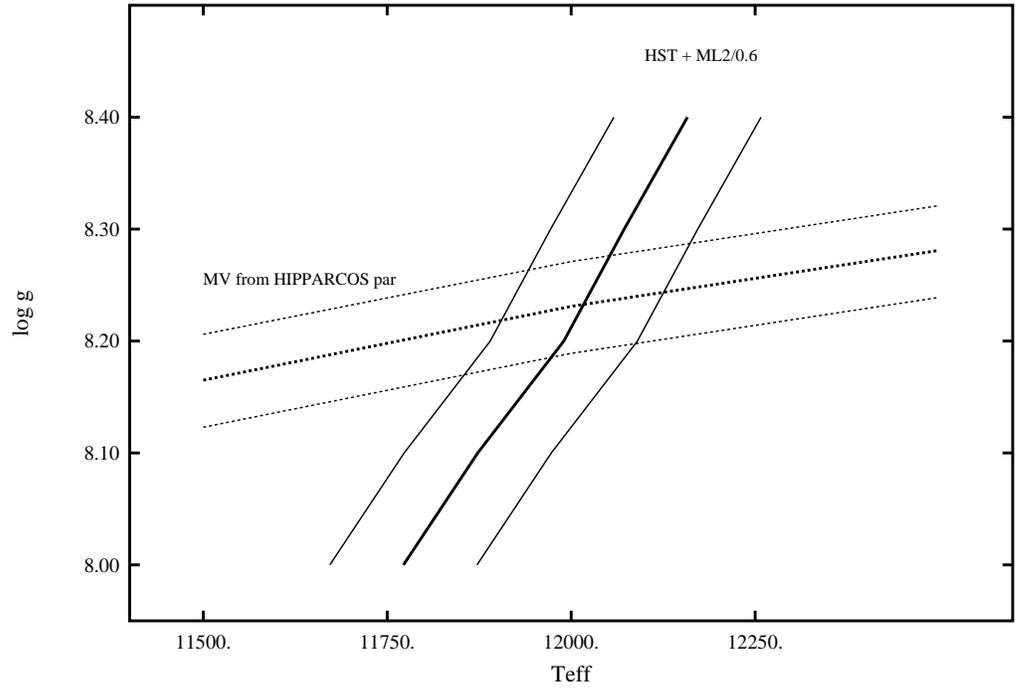}
\figcaption[fig8.ps]{Temperature and log g determination for G226-29 using the spectra
itself and HIPPARCOS parallax. The heavy line shows the determination
of $T_{\mathrm{eff}}$ and $\log g$ from the spectra, and their
uncertainties. The dotted lines the constraints imposed by the HIPPARCOS
parallax.
\label{koester}}
\end{figure}

\begin{figure} 
\plotone{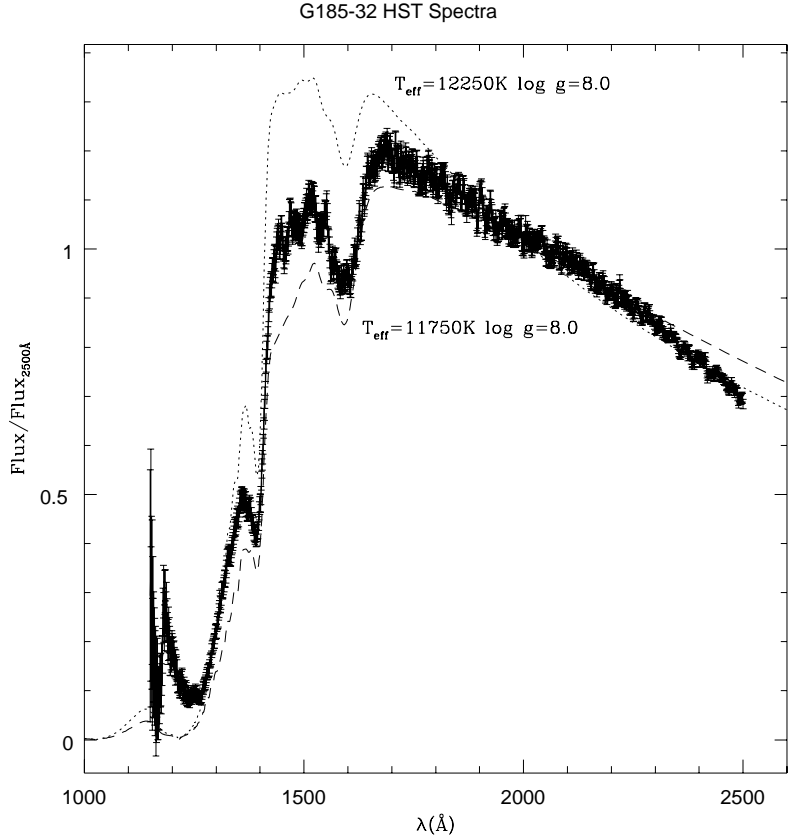}
\figcaption[fig9.ps]{Ultraviolet spectra for G185--32, with 1 $\sigma$ error-bars.
For displaying purposes, we have normalized the spectra at 2000\AA,
as the bump on the observed spectrum is due to inadequate flux
calibration of the FOS.
The dotted line represent model with $\log g=8.0$ and 
$T_{\mathrm{eff}}=12250$~K,
and the dashed line 
$T_{\mathrm{eff}}=11750$~K.
\label{g185s}}
\end{figure}

\begin{figure}
\plotone{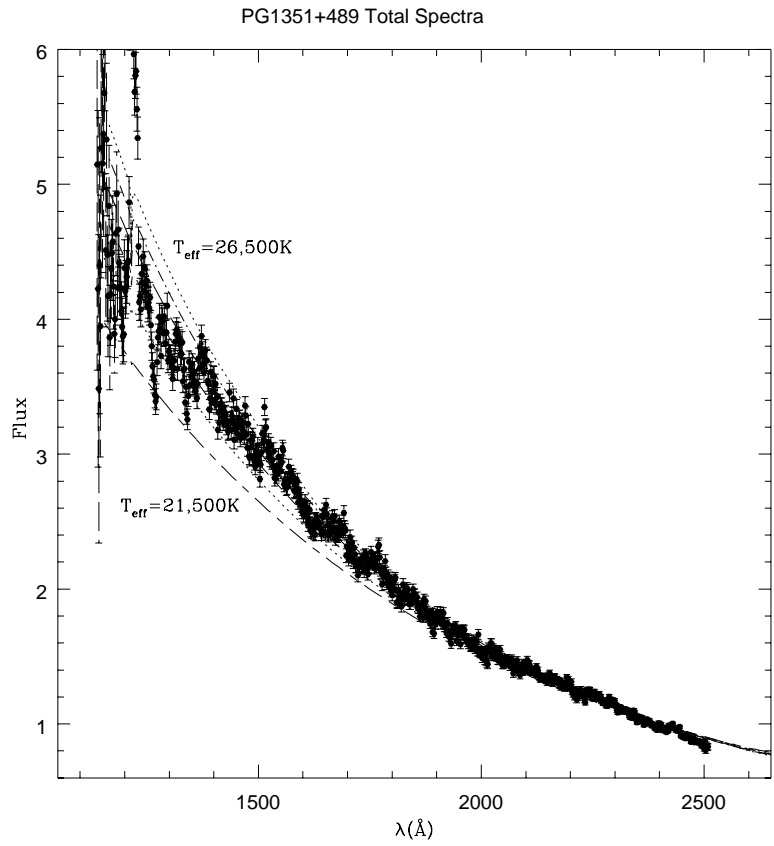}
\figcaption[fig10.ps]{Ultraviolet spectra for PG1351+489.
The lines represent models with fixed $\log g=8.0$ and different temperatures.
\label{pg1351s}}
\end{figure}

\begin{figure} 
\plotone{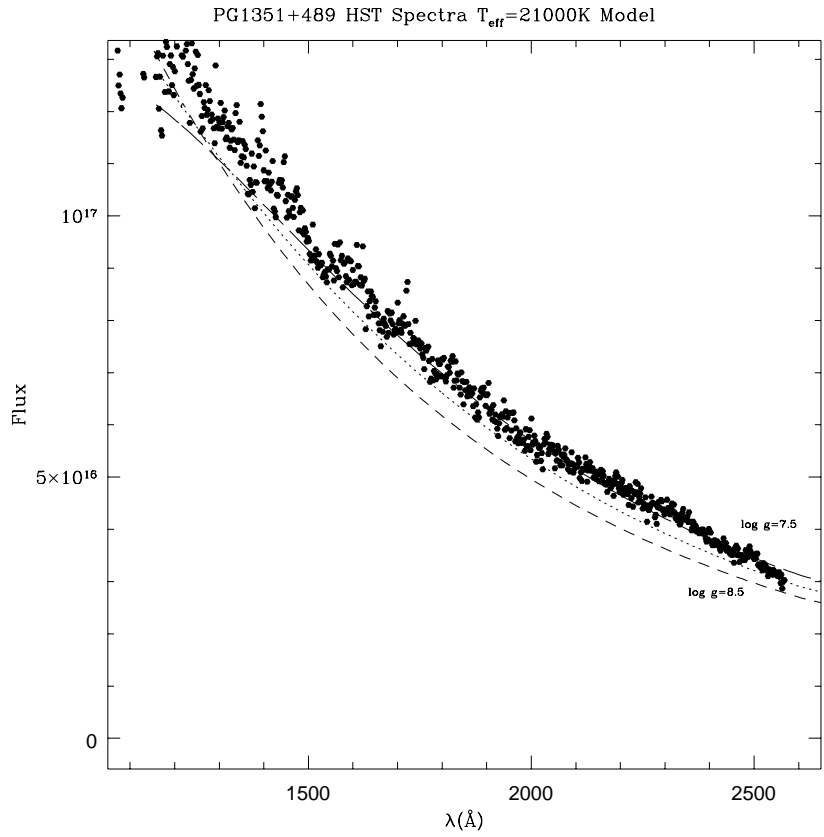}
\figcaption[fig11.ps]{Flux change due to changes in $\log g$ for DB models
showing the spectra itself is not very sensitive to surface
gravity.
\label{pg1351s1}}
\end{figure}

\begin{figure} 
\epsscale{0.8}
\plotone{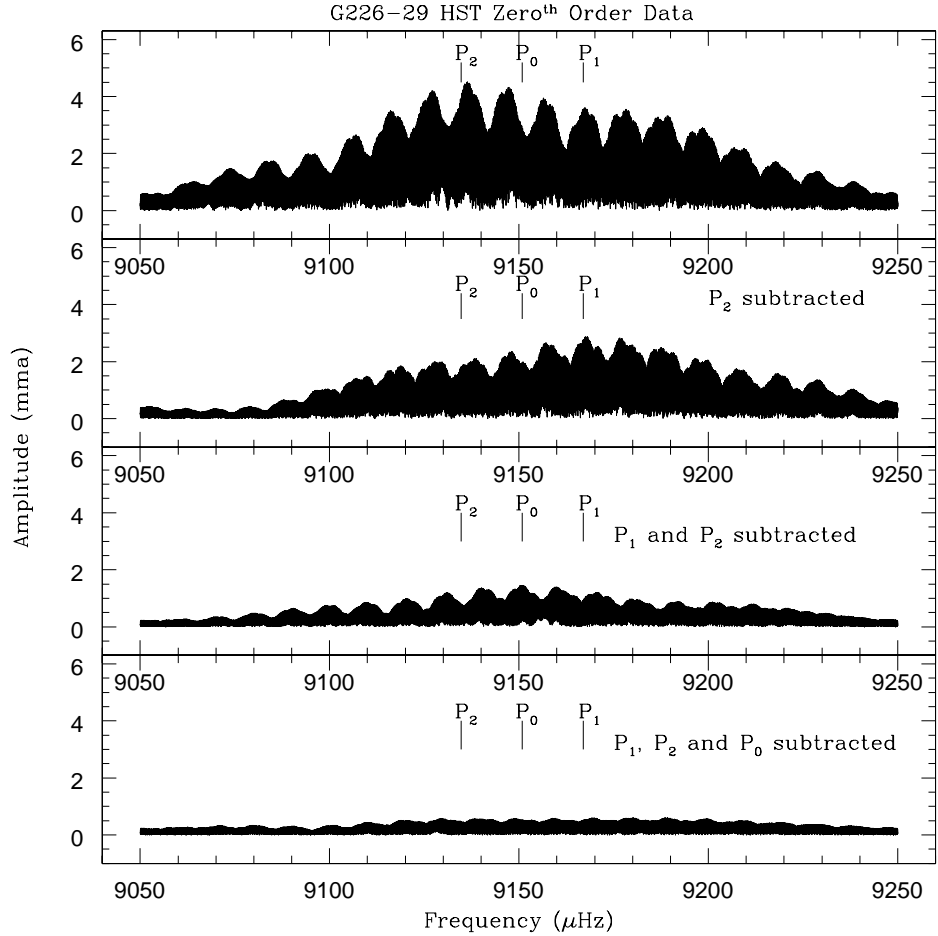}
\figcaption[fig12.ps]{Fourier amplitude spectra of the HST data set. 
The second panel shows the amplitude spectra after
we subtracted the periodicity $P_2$ from the light curve 
(prewhitening). The
lower panel shows
the amplitude spectra after
we subtracted from the light curve all three periodicities $P_1$, $P_2$
and $P_2$.
The leftover power corresponds to $P_0$.
\label{Fig4}}
\end{figure}

\begin{figure} 
\plotone{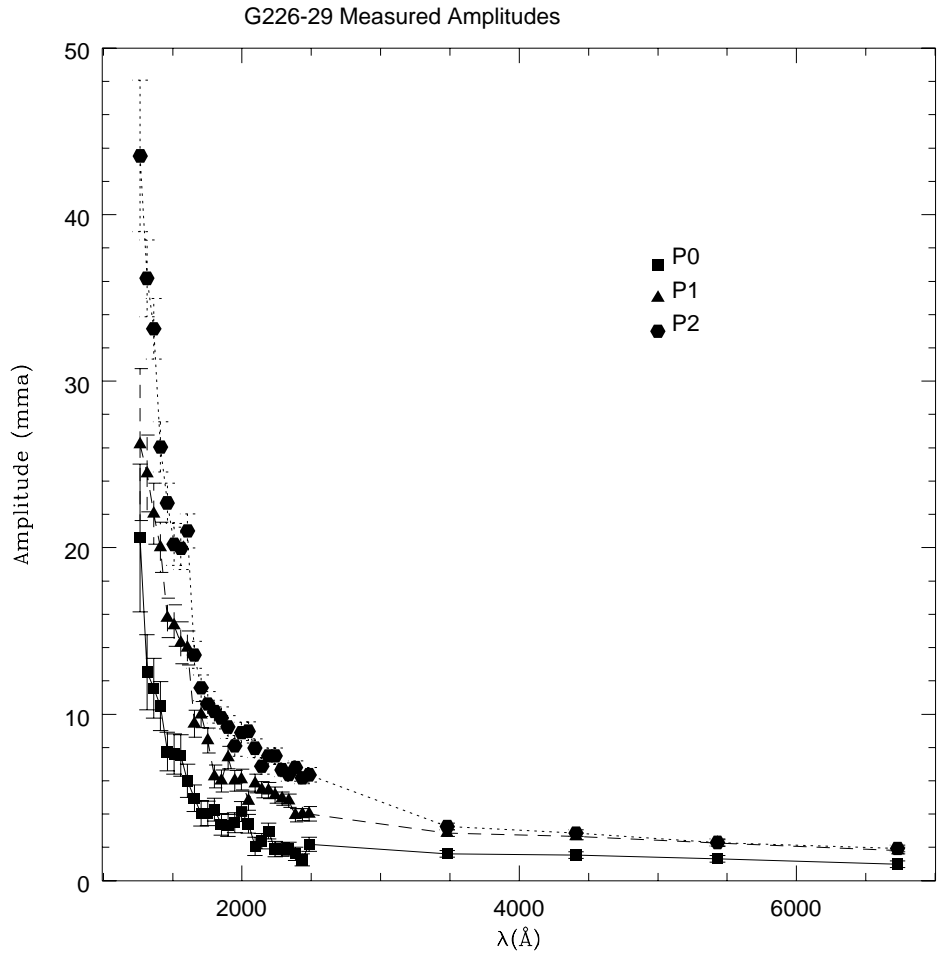}
\figcaption[fig13.ps]{Measured Amplitudes for the three modes of G226-29.
The lines simply connect the observations, to guide the eye.
\label{Fig5}}
\end{figure}

The original HST data set consists of 
764 useful pixels from 1180~\AA\ to 2508~\AA, each 
with a width of 1.74~\AA,
but we can only measure reliable amplitudes for the bins
redder than 1266~\AA.
We convolved the theoretical amplitude spectra (Figure \ref{Fig1})
with
the measurements summed into 50~\AA\ bins, over the ultraviolet, and
the UBVR transmission curves for the optical, obtaining
amplitudes directly comparable to the normalized measurements
(Table 2).

\begin{figure} 
\plotone{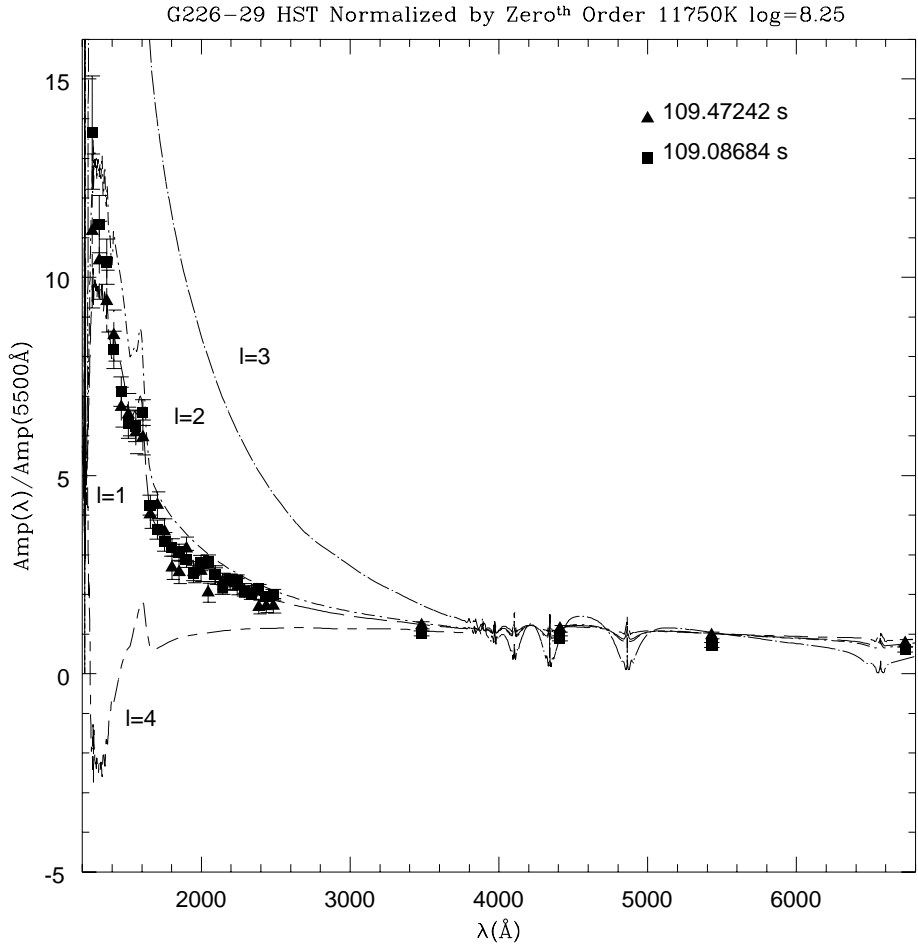}
\figcaption[fig14.ps]{The amplitudes of the $\ell = 1$ to $\ell = 4$ pulsation
modes in a pulsating DA white dwarf as a function of wavelength.
The points are the measured amplitudes of the two main periodicities
of G229-29.
\label{g226a}}
\end{figure}

We then proceeded with a least-squares fit of the amplitude vs.
wavelength observed curves to the theoretical ones, and for G226--29
modes $P_1$ and $P_2$ fit an $\ell=1$
g-mode, as shown in Figure~\ref{g226a}, and fail to fit the other
modes shown. The central mode, $P_0$, fits $\ell=1$ best, but the
data are too noisy to exclude $\ell=2$.

We determined the $\ell$, $T_{\mathrm{eff}}$, and $\log g$ independently
for each of three modes, $P_0$, $P_1$ and $P_2$,
by fitting 
${\mathrm{Amp}}(\lambda)/{\mathrm{Amp}}(5500{\mathrm{\AA}})$
for each periodicity to the model grid, with $\ell$,
$T_{\mathrm{eff}}$ and $\log g$ as free parameters,
by least squares.
Each mode can be fitted 
by $\ell=1$ or $\ell=2$ with different 
$T_{\mathrm{eff}}$ and $\log g$, but
all modes fit only one model, 
with $T_{\mathrm{eff}} = 11\,750 \pm 20$~K and
$\log g = 8.23 \pm 0.06$, for $\ell=1$.
Note that the quoted uncertainties are
only those of the least-squares fit and do not represent the true
uncertainties. There are substantial systematic errors introduced by the
normalization of the flux at 3400~\AA,
the HST flux calibration, as well as the uncertainties due to the
mixing-length approximation used in the model atmospheres
(Bergeron et al. 1995, Koester \& Vauclair 1997,
Koester \& Allard 2000), which is
incapable of representing the true convection in the star at
different depths (Wesemael et al. 1991, Ludwig, Jordan \& Steffan 1994).

\subsection{G185--32}
\begin{figure} 
\plotone{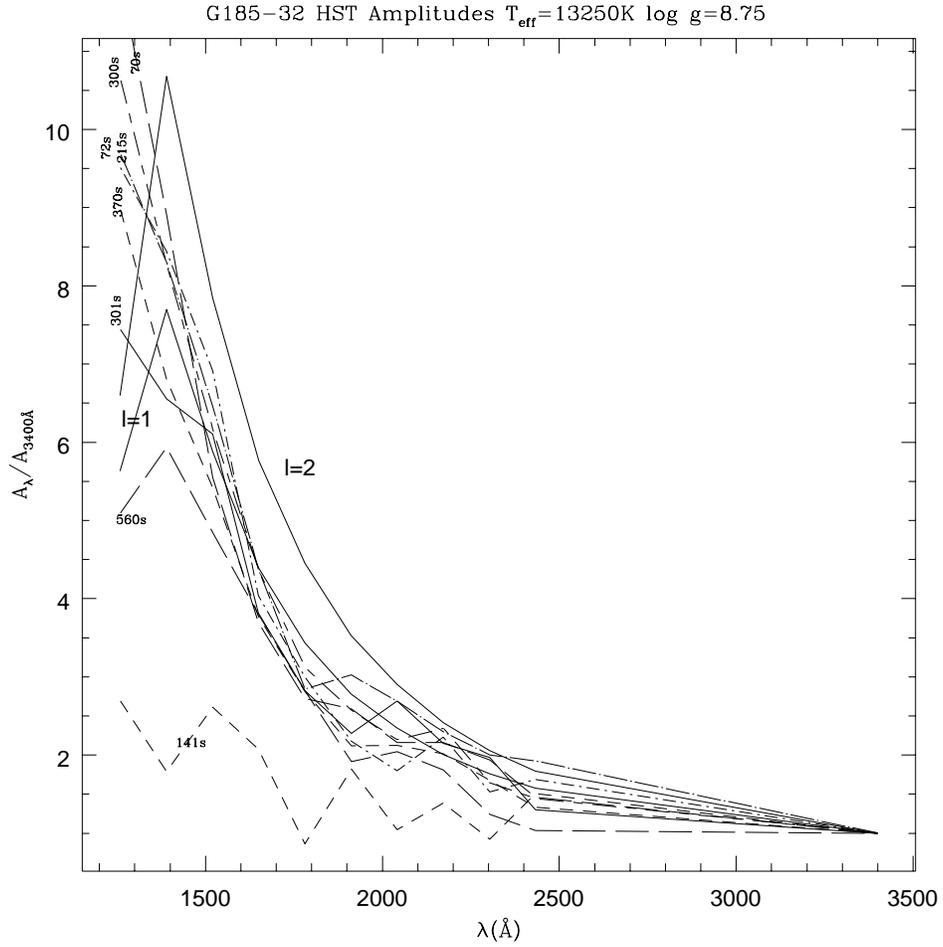}
\figcaption[fig15.ps]{Comparison of the model amplitudes (solid lines)
with the data for G185--32,
when we allow $T_{\mathrm{eff}}$, $\log g$ and $\ell$ vary.
The 141~s periodicity does not rise towards the UV, indicating
it is not caused by a g-mode pulsation.
\label{g185l2}}
\end{figure}

\begin{figure} 
\plotone{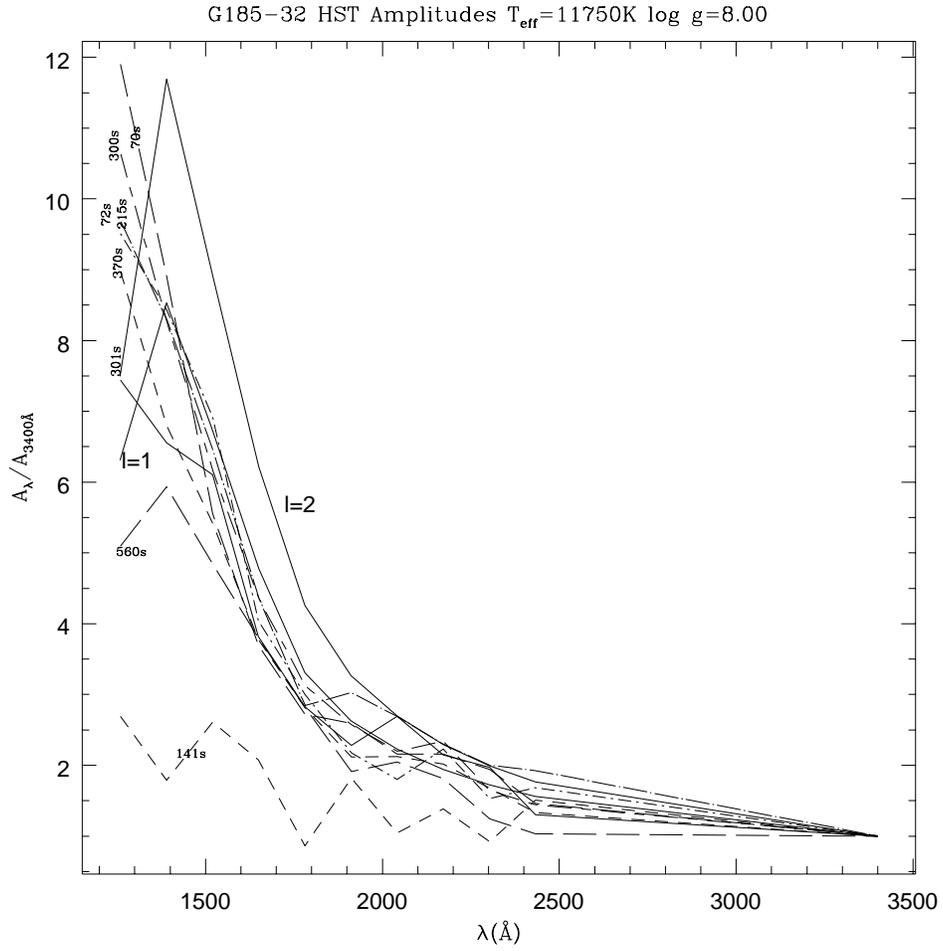}
\figcaption[fig16.ps]{Comparison of the model amplitudes 
(solid lines)
with the data for G185--32,
when we fix the $T_{\mathrm{eff}}=$11\,750~K and $\log g=8.0$,
consistent with the optical and UV spectra, V and parallax.
\label{Fig14}}
\end{figure}

For G185--32 a fit of the change in
pulsation amplitude with wavelength (Table 3)
with all three parameters: $T_{\mathrm{eff}}$,
$\log g$ and $\ell$ free resulted in $\ell$ being 
consistent with
either 1 or 2,
but the required temperatures and gravities for $\ell=2$,
$T_{\mathrm{eff}}=13\,250$~K and $\log g=8.75$ (Figure \ref{g185l2})
were inconsistent
with those derived from the time-averaged spectrum itself
$T_{\mathrm{eff}}=11\,750$~K and $\log g=8.0$. We therefore fixed
the temperature and gravity to those derived using the time-averaged spectra,
V magnitude and parallax (Koester \& Allard 2000)
and fitted the amplitude variation for $\ell$.
$\ell=1$ is the best fit for all the modes, 
except the 141~s mode of G185--32
that does not fit any pulsation index, because its
amplitude does not change significantly in the ultraviolet 
(Figure \ref{Fig14}),
in contradiction to what is expected from the
theoretical models for a DAV.
As for G226--29, and PG1351+489, our normalization uses the amplitude
of the Zeroth order data because it has the same Fourier spectral
window as the ultraviolet data.

\subsection{PG1351+489}
\begin{figure} 
\plotone{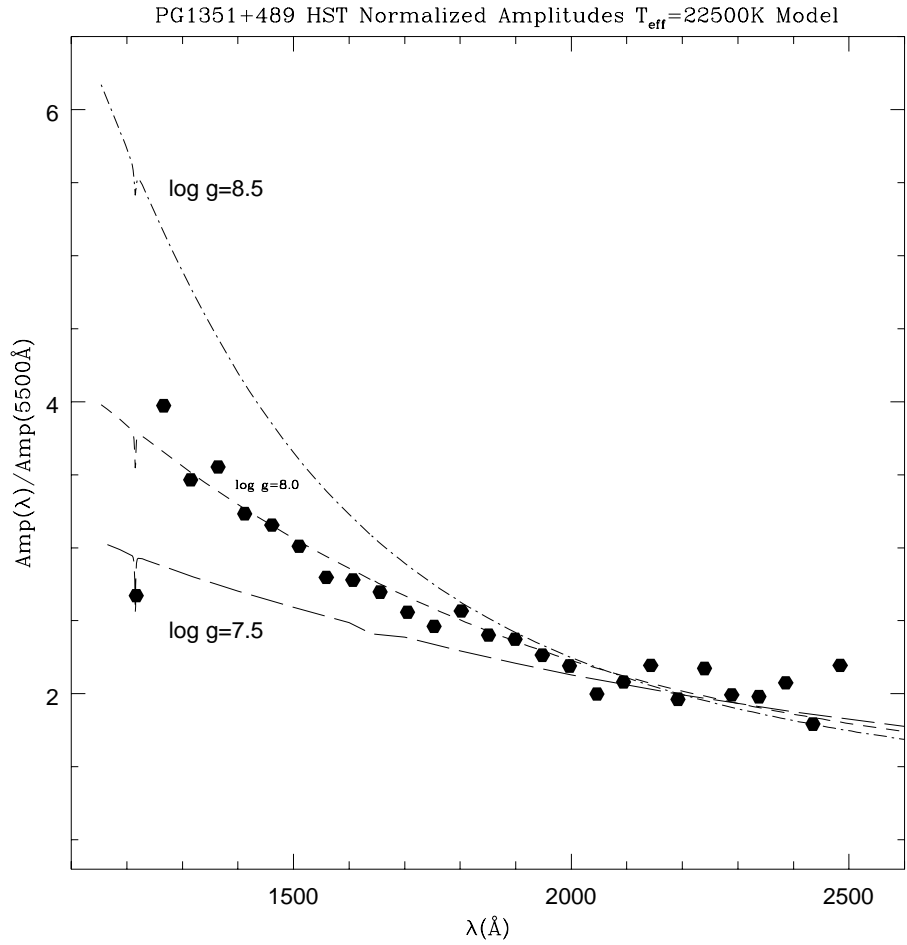}
\figcaption[fig17.ps]{Amplitude change due to changes in $\log g$ for DB models.
The lines represent 
the normalized amplitudes expected for
models with fixed $T_{\mathrm{eff}}=$22\,500~K
and different values for $\log g$.
Even though the spectra are not very sensitive to surface
gravity, the amplitudes are.
\label{Fig18}}
\end{figure}

For PG1351+489, a fit of the normalized ultraviolet
amplitudes to the theoretical ones,
for the main periodicity at 489~s and its harmonic at 245~s,
fit an $\ell=1$ mode, with
$T_{\mathrm{eff}} = 22\,500~{\mathrm{K}} \pm 250$~K, and
$\log g = 8.0 \pm 0.10$,
or an $\ell=2$ mode with $T_{\mathrm{eff}} = 23100~K \pm 250$K
and $\log g=7.5 \pm 0.10$. Figure \ref{Fig18} shows that the amplitude
ratios are dependent on $\log g$, but again the temperature and
$\log g$ determination cannot be untangled.
As the optical spectra of Beauchamp et al. (1999)
indicate $\log g = 7.9$ for PG1351+489, and $\log g \simeq 8.0$ for the
whole DBV class,
we conclude that the best solution is $\ell=1$ and 
$\log g\simeq 8.0$.

It is important to note that all pulsations have the same phase
at all wavelengths, to within the measurement error of a few seconds,
and therefore no phase shift with wavelength is detected,
assuring that all geometric and some non-adiabatic effects are negligible;
the main non-adiabatic effect is a phase shift between the motions
(velocities) and the flux variation, not measurable in our data.

\section{Discussion}

For a DAV, an $\ell=1$ mode with 109~s period requires a $k=1$ radial index
from pulsation calculations, and therefore 
the model for
G226--29 has to have
a thick hydrogen surface layer, around $10^{-4}\mstr$ (Bradley 1998).
The star may have a thick hydrogen layer as well.
The effective temperature derived from the pulsation
amplitudes, $T_{\rm{eff}}=11750$, and surface
gravity $\log g=8.23$, indicate a mass of $(0.75 \pm 0.04) \msun$,
according to the
evolutionary models of Wood (1992). 
As all three modes fit $\ell=1$, they must be a triplet from a
rotationally split $\ell=1$ mode.

Even though the 
Bergeron et al. (1995) ML2/$\alpha=0.6$ instability strip runs
$12460~{\rm K} \geq T_{\mathrm{eff}} \geq 11160~{\rm K}$, 
we know by comparison of its spectra
with other ZZ Cetis that G226--29 is at the blue edge. 
As it is at the blue
edge, such a low temperature indicates
the instability strip is at much lower temperature than previously quoted.
Also,
the observed low amplitude of the pulsations, their short period and the small
number of pulsations all indicate it is at the blue edge for its
mass, and the higher than average mass suggests a higher temperature
instability strip. According to Bradley \& Winget (1994), the
ML3 instability strip for a 0.75~$\msun$ white dwarf is at 13\,100~K,
330K hotter than for a 0.6~$\msun$ star.
Giovannini et al. (1998) show that the observed instability strip does
depend on mass, as Bradley \& Winget (1994) predicted.
One of the problems with the determination of an effective temperature
for a star is that it may vary with the wavelength used in the determination;
even though the effective temperature is a bolometric parameter,
none of our observations are. The problem is dramatic for stars
with surface convection layers, because of the effects
of turbulent pressure on the photosphere.
None of the model atmospheres calculated
with mixing length theory can reproduce the physical non-local
processes involved (Canuto \& Dubovikov 1998),
requiring different parameterizations at different depths (Ludwig,
Jordan \& Steffen 1994), and a fine tuning of the convection
mixing length coefficient for the wavelength region of
interest (Bergeron et al. 1995, Koester \& Vauclair 1997,
Koester \& Allard 2000). The model atmospheres used assume the atmosphere is
in hydrostatic and thermodynamical equilibrium, an assumption
that must be examined because the timescale for convection
is of the same order of the timescale for pulsation.

The periodicity at 141~s for G185--32 does not change its amplitude
significantly with wavelength and therefore does not fit any theoretical
model. As its period is twice that of the 70.92~s periodicity, one
must consider if it is only a pulse shape effect on the 70.92~s
periodicity, but normally a pulse shape effect occurs as an harmonic,
not a sub-harmonic, because it normally affects the rise and fall of
the pulse.  We have no plausible explanation for this periodicity,
showing that these stars still have much to teach us.

The periodicity at 560~s rises slowly to the ultraviolet, but as the
HST data sets are short, its amplitude has a large uncertainty due to
aliasing, as shown by the phase change from ultraviolet ($-1.5 \pm 8.2$~s)
to zeroth order data ($44.8 \pm 9.6$~s).

Another surprise for G185--32 is the identification of the
periodicities at 70.93~s and 72.56~s as $\ell=1$ modes. The pulsation
models are consistent with such short period $\ell=1$, k=1 mode only
for a total mass around $1~M_\odot$. Higher k values
require even larger mass. But the mass determination
from our time average spectra, as well the mass determinations
by Koester \& Allard (2000) using the IUE spectra plus V mag and parallax,
or the optical spectra of Bergeron et al. (1995) all derive a
normal mass around 0.56~$M_\odot$. One possibility to resolve such
difference is if the observed modes were a rotationally split
k=0 mode, but that would require that G185--32 be a binary star,
to account for the center of mass changes during pulsation. As
G185--32 is not a known binary star, such explanation requires the
discovery of a companion, possibly with a high signal-to-noise
red spectra.

The models, while useful, clearly lack some of the pulsation
physics present in the star.

The splitting of the 70.93 to 72.56~s periodicities, assuming they
are m-splitting of the same k and $\ell$ mode,
imply a rotation period around 26~min,
but that of the 299.95~s to 301.46~s periodicities would imply
a rotation period of 9.7~hr.
In the estimate we used
\[P_{\mathrm{rot}} = \frac{1-C_{kl}^I}{\Delta f}\]
where $\Delta f$ is the frequency splitting, and used
$C_{kl}^I\simeq 0.47$ to 0.48 for k=1, $\ell=1$. The
asymptotic value for $C_{kl}^I$ is 0.5, for $k\gg 1$.
Either value for the  rotation period is much shorter than for normal 
ZZ Ceti stars (around 1 day) so we must also consider the possibility
that the 141~s periodicity, which does not follow the g-mode theoretical
prediction, and is harmonically related to the 70.93~s periodicity,
must arise from some other cause. 
The 141~s periodicity of G185--32 is a periodic brightness change
that is not accompanied by a change in color, suggesting some kind of 
geometric effect.

\acknowledgments
We are thankful to Bob Williams, the former Director of the STScI
for granting us director's discretionary time for the project,
to Jeffrey Hayes, our project scientist at STScI, for the
continuous help with the HST data reduction and to Bill Welsh,
from the University of Texas, for bringing the zeroth order data to
our attention.
Support for this work was provided by NASA through grants
number GO-5581, GO-6011 and
GO-6442 from the Space Telescope Science Institute, which is
operated by AURA, Inc., under NASA contract NAS5-26555,
Support to S.O.K. was also provided by CNPq-Brazil. D.K.
acknowledges financial support for work on HST observations
from the DLR through grant No. 50 OR 96173.
Jiang acknowledges financial support from Chinese Natural Science Foundation, grant No. 19673008.

\clearpage

\begin{table}
\begin{center}
\caption{Optical Amplitudes for G226--29}
\begin{tabular}{cccl}
\tableline
	Color
	   & $Amp_1$
	   & $Amp_0$
	   & $Amp_2$\cr
\tableline
U  & 1.61 $\pm$ 0.09 & 2.87 $\pm$ 0.09 &  3.25 $\pm$ 0.09  \cr
B  & 1.53 $\pm$ 0.06 & 2.66 $\pm$ 0.06 &  2.86 $\pm$ 0.06  \cr
V  & 1.27 $\pm$ 0.08 & 2.30 $\pm$ 0.08 &  2.28 $\pm$ 0.08  \cr
R  & 1.00 $\pm$ 0.10 & 1.81 $\pm$ 0.10 &  1.94 $\pm$ 0.10  \cr
\tableline
\end{tabular}
\label{UBVR}
\end{center}
\end{table}

\begin{table}
\begin{center}
\tablecaption{Measured Amplitudes in mma for G226--29}
\caption{Measured Amplitudes in mma for G226--29}
\begin{tabular}{cccr}  \tableline
         $\lambda_{\mathrm{ef}}$
           & $\mathrm{Amp_0}$
           & $\mathrm{Amp_1}$
           & $\mathrm{Amp_2}$\\
\tableline
1266 & 20.58 $\pm$ 4.44 & 26.20 $\pm$ 4.56  & 43.53 $\pm$ 4.56 \\
1315 & 12.52 $\pm$ 2.25 & 24.26 $\pm$ 2.31 & 36.18 $\pm$ 2.31\\
1364 & 11.56 $\pm$ 1.79 & 22.04 $\pm$ 1.83 & 33.15 $\pm$ 1.83 \\
1412 & 10.48 $\pm$ 1.47 & 20.02 $\pm$ 1.51 & 26.05 $\pm$ 1.51 \\
1461 & 7.76 $\pm$ 1.15 & 15.78 $\pm$ 1.18 & 22.69 $\pm$ 1.18 \\
1510 & 7.61 $\pm$ 1.22 & 15.33 $\pm$ 1.25 & 20.20 $\pm$ 1.25 \\
1559 & 7.52 $\pm$ 1.24 & 14.29 $\pm$ 1.27 & 19.96 $\pm$ 1.27 \\
1607 & 6.01 $\pm$ 1.00 & 13.98 $\pm$ 1.03 & 21.01 $\pm$ 1.03 \\
1656 & 4.95 $\pm$ 0.79 & 9.42 $\pm$ 0.81 & 13.56 $\pm$ 0.81 \\
1705 & 4.04 $\pm$ 0.76 & 9.98 $\pm$ 0.78 & 11.59 $\pm$ 0.77 \\
1753 & 4.04 $\pm$ 0.73 & 8.42 $\pm$ 0.75 & 10.62 $\pm$ 0.75 \\
1802 & 4.27 $\pm$ 0.68 & 6.27 $\pm$ 0.69 & 10.16 $\pm$ 0.69 \\
1851 & 3.38 $\pm$ 0.65 & 6.00 $\pm$ 0.66 & 9.78 $\pm$ 0.66 \\
1899 & 3.33 $\pm$ 0.66 & 7.40 $\pm$ 0.68 & 9.23 $\pm$ 0.68 \\
1948 & 3.48 $\pm$ 0.60 & 6.01 $\pm$ 0.62 & 8.10 $\pm$ 0.62 \\
1997 & 4.15 $\pm$ 0.60 & 6.08 $\pm$ 0.62 & 8.90 $\pm$ 0.62 \\
2046 & 3.42 $\pm$ 0.56 & 4.80 $\pm$ 0.57 & 8.98 $\pm$ 0.57 \\
2094 & 2.06 $\pm$ 0.54 & 5.84 $\pm$ 0.55 & 7.96 $\pm$ 0.55 \\
2143 & 2.38 $\pm$ 0.46 & 5.46 $\pm$ 0.48 & 6.88 $\pm$ 0.48 \\
2192 & 2.98 $\pm$ 0.48 & 5.42 $\pm$ 0.49 & 7.49 $\pm$ 0.49 \\
2240 & 1.90 $\pm$ 0.47 & 5.14 $\pm$ 0.48 & 7.49 $\pm$ 0.48 \\
2289 & 1.87 $\pm$ 0.40 & 4.93 $\pm$ 0.41 & 6.66 $\pm$ 0.41 \\
2338 & 1.92 $\pm$ 0.39 & 4.81 $\pm$ 0.40 & 6.38 $\pm$ 0.40 \\
2386 & 1.61 $\pm$ 0.39 & 3.94 $\pm$ 0.40 & 6.80 $\pm$ 0.40 \\
2435 & 1.25 $\pm$ 0.36 & 3.98 $\pm$ 0.36 & 6.18 $\pm$ 0.36 \\
2484 & 2.19 $\pm$ 0.42 & 4.02 $\pm$ 0.43 & 6.36 $\pm$ 0.43 \\
3400 & 1.22 $\pm$ 0.32 & 2.72 $\pm$ 0.32 & 3.70 $\pm$ 0.32 \\
\tableline
\end{tabular}
\end{center}
\end{table}

\begin{table}
\begin{center}
\caption{Amplitudes for G185--32}
\begin{tabular}{ccccc}  \tableline
 & Ultraviolet & & Zeroth Order &\\ \tableline
Period (s) & Amplitude (mma) & Time of maxima (s) & Amplitude (mma) & Time of maxima (s)\\ \tableline
215.7 & 7.81 $\pm$ 0.34 & 61.7 $\pm$ 1.6 & 2.68 $\pm$ 0.15 & 61.8 $\pm$ 2.0 
\\ 
370.1 & 4.66 $\pm$ 0.36 & 106.1 $\pm$ 4.5 & 2.15 $\pm$ 0.16 & 99.6 $\pm$ 4.3 
\\ 
70.9 & 4.48 $\pm$ 0.36 & 29.8 $\pm$ 0.9 & 1.81 $\pm$ 0.17 & 28.5 $\pm$ 1.0
 \\
72.5 & 3.18 $\pm$ 0.36 & 26.6 $\pm$ 1.3 & 1.21 $\pm$ 0.16 & 21.4 $\pm$ 1.6
 \\
301.3 & 4.22 $\pm$ 0.36 & 25.1 $\pm$ 4.1 & 1.90 $\pm$ 0.16 & 289.0 $\pm$ 4.0
 \\
300.0 & 4.14 $\pm$ 0.36 & 182.8 $\pm$ 4.1 & 1.86 $\pm$ 0.16 & 199.6 $\pm$ 4.1\\ 
560.0 & 3.37 $\pm$ 0.36 & 44.8 $\pm$ 9.6 & 1.74 $\pm$ 0.16 & 558.5 $\pm$ 8.2  \\ 
141.8 & 1.85 $\pm$ 0.37 & 98.3 $\pm$ 4.5 & 1.56 $\pm$ 0.16 & 103.2 $\pm$ 2.3  \\ \tableline
\end{tabular}
\end{center}
\end{table}

\clearpage

\clearpage 

\clearpage

\clearpage

\clearpage

\clearpage

\clearpage

\clearpage

\clearpage

\clearpage 
\clearpage

\clearpage

\clearpage

\clearpage

\clearpage

\clearpage

\clearpage

\end{document}